\documentclass[amssymb,prl,aps,twocolumn,amsmath,showpacs]{revtex4}

\usepackage[dvips]{graphicx}
\usepackage{amstext}

\begin{document}

\title{Optimization of spin-triplet supercurrent in ferromagnetic Josephson junctions}
\author{Carolin Klose, Trupti S. Khaire, Yixing Wang, W. P. Pratt, Jr., Norman O. Birge}
\email{birge@pa.msu.edu}
\affiliation{Department of Physics and
Astronomy, Michigan State University, East Lansing, Michigan
48824-2320, USA}
\author{B.J. McMorran}
\author{T.P. Ginley}
\altaffiliation[Present address: ]{Physics Department, Juniata
College, Huntingdon, PA 16652, USA.}
\author{J.A. Borchers, B.J. Kirby, B.B. Maranville, J.
Unguris}
\affiliation{National Institute of Standards and
Technology, Gaithersburg, Maryland 20899, USA}

\date{\today}

\begin{abstract}

In the past year, several groups have observed evidence for
long-range spin-triplet supercurrent in Josephson junctions
containing ferromagnetic (F) materials.  In our work, the
spin-triplet pair correlations are created by non-collinear
magnetizations between a central Co/Ru/Co ``synthetic
antiferromagnet" (SAF) and two outer thin F layers. Here we
present data showing that the spin-triplet supercurrent is
enhanced up to 20 times after our samples are subject to a large
in-plane magnetizing field. This surprising result can be
explained if the Co/Ru/Co SAF undergoes a ``spin-flop" transition,
whereby the two Co layer magnetizations end up perpendicular to
the magnetizations of the two thin F layers.  Direct experimental
evidence for the spin-flop transition comes from scanning electron
microscopy with polarization analysis and from spin-polarized
neutron reflectometry.

\end{abstract}

 \maketitle

Experimental and theoretical progress in
superconducting/ferromagnetic (S/F) hybrid systems has been
impressive over the past decade \cite{BuzdinReview}.  When a
conventional spin-singlet Cooper pair crosses the S/F interface,
the two electrons enter into different spin bands in F with
different Fermi wavevectors \cite{Demler}.  The resulting
oscillations in the pair correlation function lead to oscillations
in several observable quantities \cite{BuzdinReview}, but
unfortunately the oscillations decay exponentially as soon as the
F-layer thickness exceeds the electron mean free path
\cite{Bergeret:01b}.

In contrast to spin-singlet electron pairs, spin-triplet pairs can
survive in F as long as they would in a normal metal.  While
spin-triplet superconductivity arises only rarely in bulk
materials \cite{Mackenzie, Saxena}, it was predicted a decade ago
that such pairs can be induced in S/F hybrid systems in the
presence of certain kinds of magnetic inhomogeneity involving
non-collinear magnetizations
\cite{Bergeret:01a,Kadigrobov:01,Eschrig:03}. Experimental
evidence for such spin-triplet pairs was elusive for many years
\cite{Keizer:06,Sosnin:06}; then, last year, several groups
published convincing evidence for spin-triplet supercurrent in
S/F/S Josephson junctions containing only conventional
spin-singlet S materials
\cite{Khaire:10,Robinson:10,Sprungmann:10,Anwar:10}. The
conversion from spin-singlet to spin-triplet pairs was
accomplished either by introducing magnetic inhomogeneity
artificially, or by relying on a source of inhomogeneity intrinsic
to the materials in the samples.  In our Josephson junctions, the
central F layer is in fact a Co/Ru/Co ``synthetic antiferromagnet"
(SAF) with the magnetizations of the two Co layers locked
anti-parallel to each other by a strong exchange field mediated by
the Ru layer (see Fig. 1).  We insert additional thin F' layers on
either side of the SAF; these extra layers are crucial to the
creation of spin-triplet pairs inside the junctions
\cite{Houzet:07}.

What happens when one tries to magnetize the junctions by applying
a large in-plane magnetic field?  One might expect the critical
supercurrent in the junctions to drop, since the generation of
spin-triplet pairs requires the presence of non-collinear
magnetization.  The main result of this paper is the observation
that just the opposite happens.  After magnetization, the critical
current, $I_c$, increases up to a factor 20 relative to its value
in the as-grown state.  This seemingly counter-intuitive result
can be understood by considering a unique property of the SAF:
when the large magnetizing field $\textbf{\textit{H}}_{app}$ is
applied, the Co magnetizations ``scissor" towards
$\textbf{\textit{H}}_{app}$. When the field is removed, the Co
magnetizations relax back to directions perpendicular to
$\textbf{\textit{H}}_{app}$.  This SAF ``spin-flop" transition was
predicted and first demonstrated over a decade
ago.\cite{Zhu:98,Tong:00}

\begin{figure}[tbh]
\begin{center}
\includegraphics[width=3.0in]{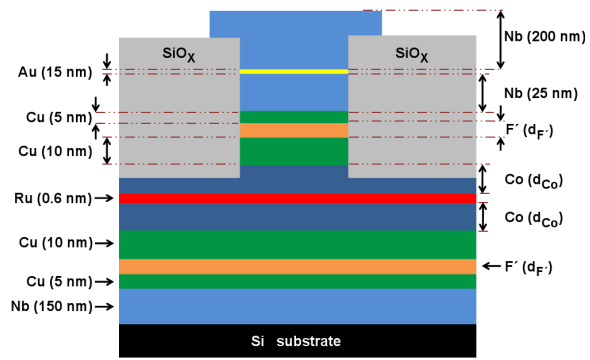}
\end{center}
\caption{(color online).  Schematic diagram of the Josephson
junctions used in the this work, shown in
cross-section.}\label{Schematic}
\end{figure}

Our sample geometry is illustrated in Fig. 1.  The two layers
labelled F' are either pure Ni or Pd$_{0.88}$Ni$_{0.12}$ alloy in
this work \cite{Khaire:10,Khasawneh:11}.  The inner Cu layers
magnetically isolate the F' layers from the Co layers.  The outer Cu
layers are present for historical reasons and because Co grows
better on a Nb/Cu buffer layer \cite{Khasawneh:09}.  The entire
multilayer except for the top Nb is sputtered in one run without
breaking vacuum. Circular junctions with diameters of 10, 20, and
40 $\mu$m are patterned by photolithography and ion milling,
followed by deposition of insulating SiO${_x}$ to isolate the top
and bottom Nb leads. Finally the top Nb electrode is sputtered
through a mechanical mask.  The purpose of the Au layer is to
suppress oxidation of the structure during processing; at low
temperature the Au becomes superconducting due to the proximity
effect with the surrounding Nb layers.  The Nb layers start to
superconduct just above 9 K; all of the data presented here were
obtained at 4.2 K.

The original purpose of the Co/Ru/Co SAF was to provide a strong
exchange field for the electrons while simultaneously producing
little to no magnetic flux in the junctions.  Large-area Josephson
junctions containing a strong ferromagnetic material such as Co
exhibit complicated and irregular ``Fraunhofer patterns" when
subject to an applied transverse magnetic field
\cite{Bourgeois:01,Khasawneh:09}. The irregularities are due to a
random pattern of constructive and destructive interference in the
gauge-invariant phase difference across the junction caused by the
complicated spatial variation of the magnetic vector potential
\cite{Khaire:09}.  The presence of the Ru restores textbook-like
Fraunhofer patterns centered very close to zero applied field
\cite{Khasawneh:09}, an indication that there is very little
intrinsic magnetic flux in the junctions. In this work, the Ru
will serve a second, unexpected role, namely to provide a simple
way to force the magnetizations of the Co layers to be
perpendicular to the magnetizations of the F' layers.

As background to the new data presented here, we briefly review
our previous results \cite{Khaire:10, Khasawneh:11}.  Josephson
junctions of the type illustrated in Fig. 1, but without the F'
layers, exhibit a critical supercurrent ($I_c$) that decays
rapidly with increasing Co thickness \cite{Khasawneh:09}. Defining
$D_{Co}$ as the sum of the thicknesses of the two Co layers, we
find that $I_c \propto exp(-D_{Co}/l_e)$ with mean free path $l_e
= 2.3$nm, for $D_{Co}$ ranging from 3 to 23 nm.  Insertion of the
F' layers with appropriate thicknesses ($d_{F'}$ between 3 and 6
nm for F'=PdNi, or $d_{F'}$ between 1 and 2 nm for F'=Ni) enhances
$I_c$ by over two orders of magnitude when $D_{Co}=20$nm.  The
dependence of $I_c$ on $D_{Co}$ is nearly flat when $D_{Co}$
varies over the range 12 - 28 nm, with F'=PdNi and $d_{PdNi}=4$nm.
This long-range behavior of the critical supercurrent is the
signature of its spin-triplet nature. For the rest of this paper
we will focus on samples containing F' layers of either PdNi or
Ni, with $D_{Co}$ fixed at 20 nm.

\begin{figure}[tbh]
\begin{center}
\includegraphics[width=3.2in]{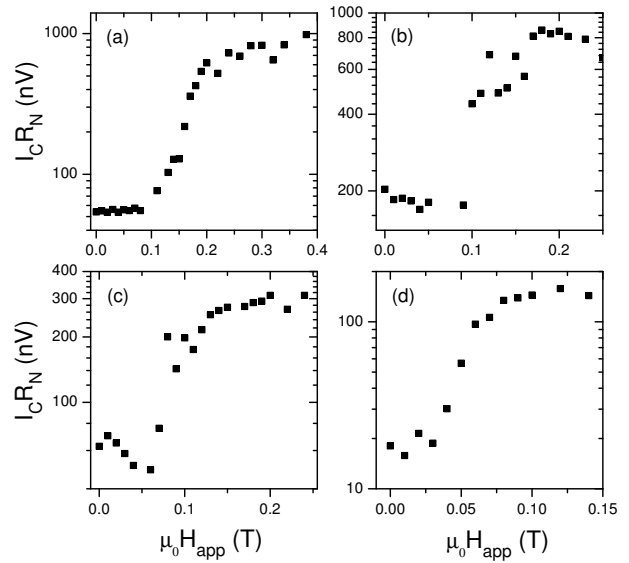}
\end{center}
\caption{Critical current times normal-state resistance ($I_cR_N$)
as a function of magnetizing field $H_{app}$ for Josephson
junctions containing F'=Ni, for Ni thicknesses $d_{Ni}=1.0, 1.5,
2.0,$ and $2.5$ nm for panels (a)-(d), respectively. Magnetizing
the samples enhances $I_c$ by a large factor that depends on
$d_{Ni}$.  (Uncertainties are dominated by changes in magnetic
configuration, and can be estimated from the scatter in the data.)
}\label{Ic_enhancement}
\end{figure}

Fig. 2 illustrates what happens when samples with F'=Ni and four
values of $d_{Ni}$ are subjected to an applied in-plane magnetic
field, $H_{app}$. After each value of field is applied, the full
Fraunhofer pattern is re-measured in the vicinity of zero field,
and we plot the maximum value of $I_c$ at the central peak of the
Fraunhofer pattern. Fig. 2 shows that, at first, very little
happens. Then when $H_{app}$ exceeds the coercive field of the Ni
layers (in the range of $\mu_0 H \approx 0.05 - 0.15$ T depending
on $d_{Ni}$), $I_c$ starts to increase dramatically.  At large
$H_{app}$, $I_c$ flattens out after having increased by a large
factor -- up to 20 for $d_{Ni}=1.0$nm. At the same time, the
central peak in the Fraunhofer patterns (not shown) shifts to a
small negative field value that is proportional to the Ni
thickness, and consistent with the remnant magnetization of our Ni
films \cite{FraunhoferShift}. The Fraunhofer shifts indicate that
the Ni layers are fully magnetized when $I_c$ saturates in Fig. 2.
Similar behavior to that shown in Fig. 2 was found in samples with
F'=PdNi with $d_{PdNi}=4$nm, but in that case the coercive field
is larger -- $I_c$ starts to increase only when $\mu_0 H_{app}$
exceeds 0.15 T, and doesn't saturate until $\mu_0 H_{app}=0.3$T.
This behavior is consistent with the very large coercive field of
PdNi thin films \cite{Khaire:09}. The Fraunhofer pattern also
shifts in the samples with F'=PdNi, but this time by a smaller
amount consistent with the remnant magnetization of our PdNi
films.

Theory predicts that the spin-triplet supercurrent in our samples
is optimized when the magnetizations of the two F' layers are
perpendicular to those of the central Co layers \cite{Houzet:07,
VolkovEfetov:10, Trifunovic:10}.  In fact, no spin-triplet pairs
should be generated when all the magnetizations in the sample are
collinear.  The large enhancement of the critical current shown in
Fig. 2 strongly suggests that magnetizing the samples optimizes
the orthogonality of the Co magnetizations with respect to the F'
magnetizations.  The small shift of the Fraunhofer pattern, on the
other hand, indicates that only the F' layers are magnetized in
the direction of $\textbf{\textit{H}}_{app}$. This scenario is
perfectly plausible in the light of the ``spin-flop" transition of
the SAF \cite{Zhu:98, Tong:00}.

\begin{figure}[tbh!]
\begin{center}
\includegraphics[width=3.2in]{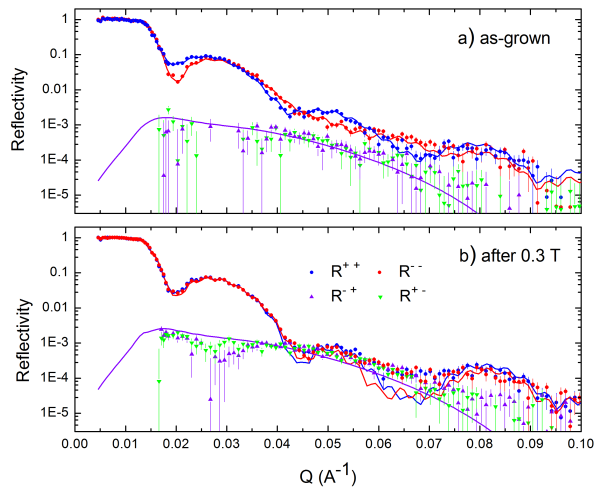}
\end{center}
\caption{(color online) Polarized neutron reflectivity data and
fits (solid lines) as a function of wavevector Q from a partial
Josephson junction in a small guide field of $<$ 0.001 T a)
as-grown and b) after application of a 0.3 T field. The
nonspin-flip cross sections, R$^{++}$ and R$^{--}$, correspond to
the blue and red circles respectively.  The spin-flip cross
sections, R$^{-+}$ and R$^{+-}$, correspond to the purple and
green triangles.} \label{fig_PNR}
\end{figure}

To identify the magnetic structure responsible for the enhancement
of the spin-triplet supercurrent, we made a large-area sample of
the form Si/Nb(150 nm)/Cu(10 nm)/Co(6 nm)/Ru(0.6nm)/Co(6nm)/Cu(10
nm), which has the Josephson junction layer structure shown in
Fig. 1 through the Cu layer on top of the upper Co layer.  The Co
magnetizations were characterized with the complementary
techniques [25] of specular polarized neutron reflectivity (PNR)
and scanning electron microscopy with polarization analysis
(SEMPA) at room temperature. (PNR measurements were also performed
at low temperature on a different sample, with results similar to
those shown here.) PNR nondestructively measures the net in-plane
magnetization for each ferromagnetic layer, even in the presence
of a field. SEMPA combined with ion milling images the remanent
magnetic structure in each layer.

The magnetization of the ferromagnetic layers was first analyzed
in the as-grown state. For the PNR measurements, performed on the
NG-1 Reflectometer at the NIST Center for Neutron Research,  the
spin states of the incident and scattered neutrons were selected
to produce the nonspin-flip (NSF) cross sections (R$^{++}$ and
R$^{--}$) and the spin-flip (SF) cross sections (R$^{+-}$ and
R$^{-+}$) shown in Figure \ref{fig_PNR}.  The NSF scattering is
sensitive to the nuclear structure of the sample, and the
splitting between R$^{++}$ and R$^{--}$ originates from the
projection of the magnetization parallel to the guide field ($<$
0.001 T).  The SF scattering is entirely magnetic and arises from
the component of the magnetization that is perpendicular to the
field.  The data were all fit with a model using the {\it Reflpak}
software package \cite{Kienzle:00} to determine the
depth-dependence of both the composition and vector magnetization
averaged across the 1 cm$^2$ area of the sample.

\begin{figure}[tbh!]
\begin{center}
\includegraphics[width=3.2in]{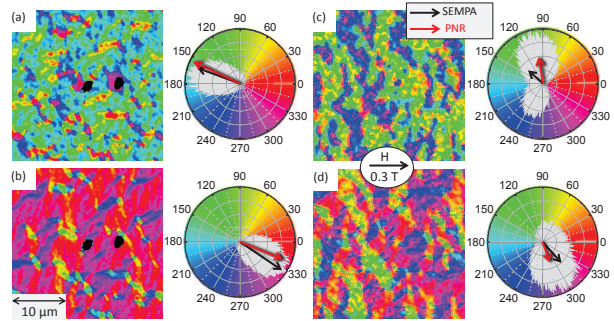}
\end{center}
\caption{(color online) SEMPA images of the magnetization in the
top (a,c) and bottom (b,d) layers before (a,b) and after (c,d) an
applied field of 0.3 T. Polar histograms to the right of each
figure show the distribution (in grey) of magnetization angles in
the image. The average magnetization from the image (black arrow)
and the magnetization measured using PNR (red arrow) are also
shown. }\label{fig_SEMPA}
\end{figure}

In Fig. \ref{fig_PNR}a), the NSF cross sections are dominated by
structural contributions, but the R$^{++}$ and R$^{--}$ exhibit a
small splitting indicative of a net moment component parallel to
the field.  The SF scattering is small, but non-zero, consistent
with a slight canting of the Co layer magnetizations away from the
field.  The red arrows in Fig. \ref{fig_SEMPA}a) and b) represent
the average orientation and magnitude of the net magnetizations of
the top and bottom Co layers obtained from the PNR fits.  SEMPA
was then used to image the magnetization of each layer within a
$\approx 1 \text{ mm}^2$ ion-milled window of the same sample. Ion
milling with 800 eV  Ar ions first reveals the top Co layer
magnetization (Fig. \ref{fig_SEMPA}a), and then the bottom Co
layer (Fig. \ref{fig_SEMPA}b). The distribution of magnetization
directions in these images are shown in the corresponding polar
plots. Both the PNR and SEMPA measurements indicate a preferred
direction for the magnetization in the as-grown state, with most
of the magnetization aligned along an angle approximately
20$^\circ$ relative to a sample edge. Both measurements also show
antiferromagnetic coupling between the top and bottom Co layers.

A 0.3 T field was then applied along the sample edge (at 0$^\circ$
in Fig. \ref{fig_SEMPA}), and the PNR and SEMPA remanent state
measurements  were repeated (Figs. \ref{fig_PNR}b,
\ref{fig_SEMPA}c and \ref{fig_SEMPA}d). The specular reflectivity
shows a pronounced increase in the SF scattering relative to the
as-grown state (Fig. \ref{fig_PNR}a) indicating that the
projections of the layer magnetizations perpendicular to the field
have increased, and the R$^{++}$  and R$^{--}$  NSF cross sections
are now essentially equal.   The red arrows in Fig.
\ref{fig_SEMPA}c) and d), obtained from the PNR fits, show that
the net magnetization of each layer has rotated in opposite
directions away from the applied field, consistent with a spin
flop transition. The PNR analysis also indicates that the
magnitudes of the Co net layer magnetizations are reduced from
their as-grown values, indicative of a decrease in average domain
size. These measurements complement the SEMPA measurements
performed on another 1 mm$^2$ area of the sample. The SEMPA images
in Figs. \ref{fig_SEMPA}c) and \ref{fig_SEMPA}d) show that the
field induces a more complicated remanent magnetic structure, with
a bimodal domain distribution within each layer that is tilted
away from the applied field. These SEMPA data are consistent with
a relaxed scissor state induced by a spin-flop transition.  A
comparison between top and bottom Co layers in both the SEMPA and
PNR results shows that they are still antiferromagnetically
coupled. (The PNR data show no direct evidence of the bimodal
distribution because the reflectivity the Co-bilayer spin
configuration shown in Figs. \ref{fig_SEMPA}c and \ref{fig_SEMPA}d
is nearly identical to that for the configuration that is mirror
symmetric about the field.  The actual spin state is clearly a
linear combination of these two.)  Also, SEMPA and PNR
measurements on similar samples containing F' layers of Ni or PdNi
confirm that the remanent magnetizations of those layers point in
the direction of $\textbf{\textit{H}}_{app}$ after application of
0.3 T.

The PNR and SEMPA measurements reveal that the evolution of the
magnetic structures within the SAF is complex, but is consistent
with a spin-flop transition. While this transition explains the
field-induced spin-triplet supercurrent enhancement, it is notable
that the state has multiple in-plane domains and the Co layer
magnetizations are tilted slightly from the direction
perpendicular to the applied field.

In conclusion, we have observed a large enhancement of the
spin-triplet supercurrent in S/F'/SAF/F'/S Josephson junctions
when the F' layers are magnetized by an applied field and the SAF
undergoes a spin-flop transition. This result confirms the
theoretical prediction that the spin-triplet supercurrent is
maximum when the magnetizations of the adjacent ferromagnetic
layers inside the junctions are aligned perpendicular to each
other. This result also underscores the need for characterization
and control of the magnetic structure to optimize the performance
of spin-triplet S/F/S devices.  This could be done in the future
by a number of methods, e.g. by exploiting shape anisotropy or by
using magnetic materials with perpendicular-to-plane anisotropy.

Acknowledgments:  We acknowledge helpful conversations with Mark
Stiles and Bob McMichael. We also thank R. Loloee and B. Bi for
technical assistance, and use of the W.M. Keck Microfabrication
Facility. This work was supported by the U.S. Department of Energy
under grant DE-FG02-06ER46341.

\end{document}